\documentclass[10pt,conference]{IEEEtran}
\usepackage[english]{babel}
\usepackage{fancyhdr}
\usepackage{framed}
\usepackage{amssymb}
\usepackage{latexsym}
\usepackage{amsmath}
\usepackage{psfrag}
\usepackage[T1]{fontenc}

\usepackage{epsfig}
\usepackage{graphicx}
\usepackage{color}

\long\def\comment#1{}

\newfont{\bbb}{msbm10 scaled 700}

\newfont{\bb}{msbm10 scaled 1100}

\newcommand{\RR}{\mbox{\bb R}}

\newcommand{\EE}{\mbox{\bb E}}

\newcommand{\xv}{{\bf x}}

\newcommand{\Sm}{{\bf S}}

\newcommand{\Um}{{\bf U}}

\newcommand{\Xm}{{\bf X}}
\newcommand{\Ym}{{\bf Y}}

\newcommand{\Nc}{{\cal N}}

\newcommand{\Qc}{{\cal Q}}
\newcommand{\Rc}{{\cal R}}

\newcommand{\Cu}{{\underline C}}

\newcommand{\SNR}{{\sf SNR}}

\newcommand{\eqdef}{\stackrel{\Delta}{=}}

\renewcommand{\SNR}{{\sf SNR}}

\begin{document}

\title{Joint Source-Channel Coding at the Application Layer for Parallel Gaussian Sources}

\author{
\authorblockN{Ozgun Y. Bursalioglu, Giuseppe Caire}
\authorblockA{Ming Hsieh Department of Electrical Engineering\\
University of Southern California \\
%Los Angeles, CA 90089, USA\\
Email: \{bursalio, caire\}@usc.edu}
\and
\authorblockN{Maria Fresia, H. Vincent Poor}
\authorblockA{Department of Electronic Electrical Engineering\\
Princeton University\\
%Princeton, NJ 08544, USA\\
Email: \{mfresia, poor\}@princeton.edu}
\thanks{This research was supported in part by the National Science Foundation under Grant CNS-06-25637.}}

\maketitle

\begin{abstract}

In this paper the multicasting of independent parallel Gaussian sources over a binary erasure broadcasted channel is considered.
Multiresolution embedded quantizer and layered joint source-channel coding schemes are used in order to serve simultaneously several users at different channel capacities.
The convex nature of the rate-distortion function, computed by means of reverse water-filling, allows us to solve relevant convex optimization problems corresponding to different performance criteria.
Then, layered joint source-channel codes  are constructed based on the concatenation of embedded scalar quantizers with binary rateless encoders.

\end{abstract}

%%%%%%%%%%%%%%%%%%%%%%%%%%%%%%%%%%%%%%%%%%%%%%%%%%%%%%
\section{Introduction} \label{sec:intro}

Multimedia streaming over heterogeneous digital networks is one of the fastest growing applications
both in terms of traffic demands and in terms of market potential.
A conventional approach consists of establishing individual streaming sessions from the server to
each users. This may be very inefficient when many users wish to receive the same content
(e.g., in mobile television or video on demand applications).
A definite advantage of analog broadcasting systems is sending simultaneously the same signal to a potentially unlimited number of receivers, with possibly different reconstruction
quality that depends on the channel conditions.
Analog transmission finds its theoretical justification in the fact that a Gaussian source under the quadratic distortion measure is
``matched'' to an Additive White Gaussian Noise (AWGN) channel \cite{GastparRV03}. Unfortunately, this lucky ``matching'' condition does not generally apply to a heterogeneous digital network (e.g., the Internet). Such a network, which may be formed by wireline backbone infrastructure combined perhaps with wireless component to the end users,  is characterized by a large variability of the ``capacity'' from the media server to the individual users. The protocol stack is basically fixed. Therefore multimedia streaming applications
are typically designed as ``overlay'' systems, that is, they act at the application layer by using a fixed transport mechanism
already built-in to the network. In this paper, following \cite{Urbanke97packetizingfor}, \cite{DCC-09}, \cite{Dana06}, we model the transport mechanism as a binary erasure broadcast channel (BEBC).
This model captures the essential behavior of a network in which the server sends a stream of packets without end-to-end flow control
(e.g., using User Datagram Protocol, UDP \cite{Kurose_Ross}) and these may be lost
because of errors or buffer overflows in the network. The receivers have perfect knowledge of packet losses since packets contain
sequence numbers, and therefore can treat missing packets as {\em erasures}.

Motivated by the above scenario, we focus on efficient Joint Source-Channel Coding (JSCC) for the transmission of
a source over the BEBC. We refer to this approach as JSCC ``at the application layer'' since we disregard the underlying physical channels, in contrast to most JSCC approaches presented in the literature (e.g., \cite{Mittal-Phamdo} and references therein).
In this paper we extend our approach for JSCC of an independent and identically distributed (i.i.d.) Gaussian source over the BEBC proposed in \cite{DCC-09}
to the case of  {\em parallel Gaussian sources}. The parallel Gaussian sources model is a good first-order approximation of the output to a linear transform operating a subband decomposition of some natural source.  For example,
the JPEG2000 image coder \cite{TaMa}  transforms the original image by using a discrete wavelet transform (DWT) and the first-order marginal statistics of the transform coefficients in the different subbands may be approximated by Gaussian random variables with different variances.

%The contribution of this paper is two-fold. In the first part, we formulate relevant optimization problems in order to optimize the
%broadcasting system using  information theoretic limits (i.e., assuming an ideal rate-distortion achieving multi-resolution source code
%for the parallel Gaussian source and an ideal broadcast channel code for the BEBC).
%Then, we design simple JSCC schemes that approach the theoretical limits in practice, with finite block length and complexity.
%%%%%%%%%%%%%%%%%%%%%%%%%%%%%%%%%%%%%%%%%%%%%%%%%%%%%%%%%%%%%
Building on our previous works \cite{DCC-09} and \cite{Ozgun-Maria-JSCC-08}, we use multi-resolution (embedded) quantization
and layered JSCC.
The allocation of source layers to different users can be optimized according to different
criteria, as discussed in Section \ref{sec:optimization}. Performing close to the theoretical limits requires coding rate adaptation
with very fine granularity. This is accomplished by using ``rateless'' raptor encoders~\cite{shokr2006IT}, which are able to produce any arbitrary number of coded symbols with a single, low complexity, encoding machine.
The key idea of our practical JSCC scheme is that (see \cite{FresiaC06} and \cite{IT_Fresia_Caire})
the rate-distortion (R-D) limit for a smooth source over a symmetric channel can be closely approached
by using scalar quantization followed by a linear encoding function that maps directly the redundant quantization indices
into channel input symbols. In the limit of arbitrarily large block length and unlimited coding/decoding complexity,
this approach provably achieves the quantizer distortion $D_{\Qc}$ at bandwidth ratio $H/C$ (channel uses per source sample)
where $C$ denotes the channel capacity and $H$ denotes the entropy rate of the quantization indices,
modeled as a discrete source. For a well-behaved source, the rate $H$
achieved by entropy-coded quantization is close to the optimal rate $\Rc(D_{\Qc})$ (where $\Rc(\cdot)$ denotes the R-D function of the source) \cite{Gish-Pierce}, \cite{Ziv}. Hence, linear encoding of quantization indices performs close to optimal.
The main advantage of the scheme, though, appears at finite block length and low encoding/decoding complexity.
In fact, our approach eliminates the need for an explicit entropy coding stage after quantization.
It is well-known that standard entropy coding is ill-conditioned with respect to residual channel decoding errors.
In contrast, our scheme reconstructs directly the  quantization symbols using a ``soft-bit'' approach, i.e., using the estimated posterior log-likelihood ratios (LLRs) generated by  a Belief Propagation (BP) iterative decoder that incorporates the a priori statistics of the source (details can be found in \cite{FresiaC06}, \cite{IT_Fresia_Caire}, \cite{Ozgun-Maria-JSCC-08} and \cite{DCC-09}).
In this way, the ``catastrophic'' behavior of entropy-coded quantization is greatly reduced.

As noted above, we implement linear encoding by using raptor codes \cite{shokr2006IT}. Recently, raptor codes have been standardized as application
layer Forward Error Correction (FEC) coding for Multimedia Broadcast/Multicast Services (MBMS) within 3GPP ~\cite{3GPP-Luby}.
We hasten to say that our application is very different from this standard.
In  3GPP ~\cite{3GPP-Luby},  the ``static broadcasting''~\cite{Shulman-Feder-StaticBC} of a common content
file is considered, where each user gathers channel observations until it has received non-erased symbols
such that the whole file can be perfectly decoded. Users may have different decoding delays depending on their erasure
probabilities.
On the contrary, in this work we consider a truly ``real-time'' multicasting where the decoding delay
is the same for all users, but each user reconstructs the source at a possibly different distortion level,
depending on its own channel capacity.
Our scheme can be naturally applied to static broadcasting (e.g., video on demand applications)
by using it in conjunction with well-known protocols such as {\em harmonic broadcasting} \cite{original-harmonic, paris-harmonic}. However, we do not investigate the details of this application in this paper.
The numerical results of Section \ref{sec:Results} show that the proposed
coding scheme can achieve end-to-end distortion performance very close to the
theoretical limits with finite block length and low encoding/decoding complexity.

%%%%%%%%%%%%%%%%%%%%%%%%%%%%%%%%%%%%%%%%%%%%%%%%%%%%
\section{BEBC and Parallel Gaussian Sources} \label{sec:bebc}

The Binary Erasure Broadcast Channel (BEBC) has input alphabet $\{0,1\}$, output alphabet $\{0,1,e\}$,
(``$e$'' denoting erasure), and is defined  by $L$ channel transition probabilities
$P^{(l)}(y|x) = 1 - \epsilon_l$ for $y = x$, $P^{(l)}(e|x) = \epsilon_l$
and $P^{(l)}(y|x) = 0$ for $y \neq x,e$.
Without loss of generality, we let  $\epsilon_1\geq\ldots\geq\epsilon_L$ and denote by
$C_l = 1 - \epsilon_l$ the capacity of the $l$-th Binary Erasure Channel (BEC).
This channel serves as a simple model for a multicast network, in which an arbitrarily large number of  users are grouped into $L$ classes, each characterized by a different channel capacity. For simplicity, we shall refer to each class as a ``user'' since, in  a multicast scenario, all users belonging to the same class are indistinguishable and achieve the same performance.

It is well-known that, under mild conditions on the erasure statistics
%\footnote{As a matter of fact, the capacity region of a broadcast channel depends only on the marginal transition probabilities~\cite{CoTh}, and hence erasures can be arbitrarily correlated.  Furthermore, erasures need not be i.i.d. in time; any erasure process such that the fraction of erased symbols converges almost surely to $\epsilon_l$ for each $l = 1,\ldots,L$ yields the same capacity region \cite{Te-Sun-Han}.}
the capacity region of the BEBC is given by~\cite{Urbanke97packetizingfor}
\begin{equation} \label{cap-region}
{\cal C} = {\Big\{}(R_1,\ldots,R_L): R_l\geq 0,\; \sum_{l=1}^L\frac{R_l}{C_l}\leq1,\, \,l\in\{1,\ldots,L\}\Big{\}}.
\end{equation}
This region is achieved by time-sharing between the vertices of the region's dominant face, defined by the
hyperplane $\sum_{l=1}^L \frac{R_l}{C_l} = 1$.
% are the points $\cv_l = (0, \ldots, C_l, \ldots,0)$.
The BEBC belongs to the class of {\em stochastically degraded broadcast channels}~\cite{CoTh}.
This implies that any message to user $l$ can be also decoded by all users $j > l$ (i.e.,
by the users with better channels).

The considered source model consists of  $s$ independent Gaussian source ``components''.
A source block of length $k$ is denoted by  $\Sm\in\RR^{s\times k}$, where its $i^{\rm th}$
row, denoted by $\Sm_{(i,:)}$, has i.i.d. elements $S_{(i,j)}\sim{\cal N}(0,\sigma_i^2)$, where we assume
$\sigma_1^2 \geq \cdots \geq \sigma^2_s$. Note that $\Sm$ contains $s\times k$ source symbols.

The encoder maps $\Sm$ into a channel input codeword $\Xm = f^{(k)}(\Sm)$,
where $f^{(k)} : \RR^{s\times k} \rightarrow \{0,1\}^{n}$ is the encoding function. At each $l$-th user decoder,
the received channel output $\Ym_l$ is mapped into a reconstructed source array
$\widehat{\Sm}_l = g^{(k)}_l(\Ym_l)$, where $g^{(k)}_l : \{0,1,e\}^{ n} \rightarrow \RR^{s\times k}$ is a decoding function.
The Mean Square Error (MSE) distortion for the $l^{\rm th}$ decoder
and the $i^{\rm th}$ source component is $D^{(k)}_{i,l} = \small{\frac{1}{k} }\EE[ \|\Sm_{(i,:)} - \widehat{\Sm}_{(i,:),l}\|^2]$,
where the expectation is with respect to the joint $k$-dimensional probability distribution of $(\Sm_{(i,:)},\widehat{\Sm}_{(i,:),l})$ induced by the source, by the channel erasures and by the coding scheme.
The average total distortion at the $l^{\rm th}$ decoder is given by
$D^{(k)}_l = \small{\frac{1}{s}}\sum_{i=1}^s D^{(k)}_{i,l}$.

The R-D function of $\Sm$ is given by the ``reverse waterfilling'' parametric form~\cite{CoTh}:
 \begin{equation}\nonumber
 {\cal R}(D)=\frac{1}{s}\sum_{i=1}^s\frac{1}{2}\max\Big\{0,\log_2\frac{\sigma_i^2}{\gamma}\Big\},\;\; D=\frac{1}{s}\sum_{i=1}^s\min\{\gamma,\sigma_i^2\}.
 \end{equation}
For each pair $(R,D)$ there exist a pair $(\gamma,m)$ such that $D = \small{\frac{1}{s}}\{\gamma m + \sum_{j=m+1}^s\sigma_{j}^2\}$,
and $R=\small{\frac{1}{s}}\sum_{j=1}^m\frac{1}{2}\log_2(\sigma_j^2/\gamma)$,
with $\gamma <\sigma_m^2$.
Similar to a single Gaussian source~\cite{Equitz-Cover-SF}, it is easy to see that parallel Gaussian sources
are also successively refinable under MSE distortion.
%%%%%%%%%%%%%%%%%%%%%%%%%%%%%%%%%%%%%%%%%%%%%%%%%%%%%%
In particular, letting $(R_1,D_1)$  and $(R_2, D_2)$ denote two points on the R-D function with $D_2 < D_1$, and letting
$(m_1,\gamma_1)$ and $(m_2,\gamma_2)$ the corresponding parameters of the reverse waterfilling formula,
the rate increment for each $i$-th source component is given by
$\max\{0,{\small \frac{1}{2s}} \log_2(\sigma_i^2/\gamma_2)\} - \max\{0,{\small \frac{1}{2s}}\log_2(\sigma_i^2/\gamma_1)\}$, corresponding
to a total rate increment per source symbol
\begin{equation} \label{rate-increment}
\Delta_R\eqdef\frac{1}{s} \left[m_1\left(\frac{1}{2}\log_2\frac{\gamma_1}{\gamma_2}\right)+\sum_{i=m_1+1}^{m_2}\frac{1}{2}\log_2\frac{\sigma_i^2}{\gamma_2}\right],
\end{equation}
where the first term corresponds to the extra quantization bits for the $m_1$ source components
that are (coarsely) quantized to achieve distortion $D_1$, and the second term corresponds to the additional quantization bits
to quantize additional $m_2 - m_1$ source components that are discarded at distortion $D_1$, but need to be encoded to achieve distortion
$D_2$.

\begin{figure*}[tpb]
\centering{\includegraphics[width=13cm,height=4.5cm]{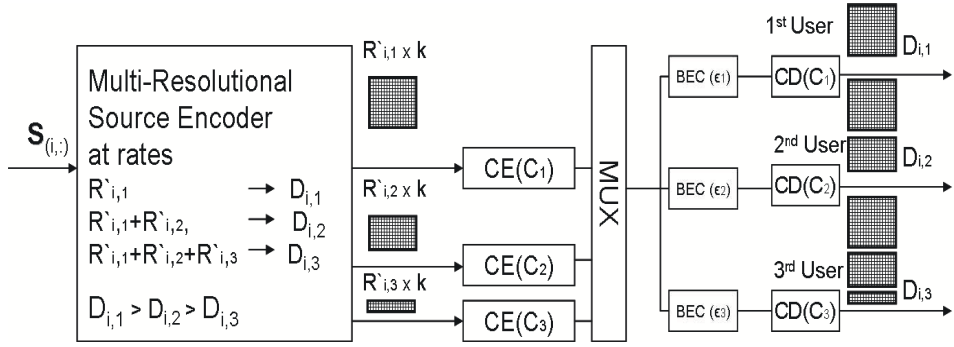}}
\caption{ A multiresolution source encoder creates 3 layers with $D_{i,1} > D_{i,2} > D_{i,3}$ for three
classes of users with capacities $C_1\leq C_2\leq C_3$.
The channel encoder-decoder pair at rate $C$ is shown as blocks $\rm{CE}(C)$ and $\rm{CD}(C)$, respectively. } \vspace{-3 mm}
\label{fig:separated}
\end{figure*}

An ideal (R-D achieving) multiresolution source code produces $L$ layers for each source component
at rates $R'_{i,l}$, for $i = 1, \ldots, s$ and $l = 1,\ldots, L$ (some layers may have zero rate).
The $L$ layers are encoded by a broadcast code for the BEBC at rates ${\underline R} = (R_1,\ldots,R_L)\in {\cal C}$.
Fig. \ref{fig:separated} shows the multiresolution layered coding scheme for the $i^{\rm th}$ component.

Inspired by the bit-per-pixel convention in image compression literature,
%\footnote{As discussed in Section \ref{sec:intro}, parallel Gaussian sources are a simple model for subbands
%of natural images produced by a DWT transform.
%In JPEG2000, the subbands have different lengths which are multiples of the length of the shortest subband,
%and their summation is equal to the number of pixels in the original image.
%In our case this corresponds to $k s$ since we have $s$ subbands with block length $k$.
%A long subband block can be divided into equal length segments, the same size of the shortest subband.
%Hence overall subbands can be modeled to have equal lengths.
%Also since we use binary erasure channels, number of channel uses corresponds to number of bits used to transmit.},
we define bandwidth expansion factor $b$ as the ratio of the number of channel uses to the number of
source symbols, i.e.  $b = n/(k s)$.
For a binary-input channel such as the BEBC, $b$ corresponds precisely to the ratio of (source-channel) coded bits per
source symbol.  Then for each layer $l$, we have $\small{\frac{1}{s}}\sum_{i=1}^s R'_{i,l} = b R_l$ and hence user $l$ achieves
the distortion
\begin{equation} \label{distortion-rate}
D_l={\cal R}^{-1}\left(\frac{1}{s}\sum_{j=1}^l\sum_{i=1}^s R'_{i,l}\right) = {\cal R}^{-1}\left(b \sum_{j=1}^l R_j\right).
\vspace{-2 mm}
\end{equation}
%%%%%%%%%%%%%%%%%%%%%%%%%%%%%%%%%%%%%%%%%%%%%%%%%%%%%%%%%%
\section{Optimization Problems} \label{sec:optimization}

A layered source-channel coding scheme can be optimized according to various criteria. Here we propose the following alternatives:

{\bf Min Bandwidth} (MB): For given target distortions $d_1 \geq \cdots \geq d_L$, we wish to
minimize $b$. Using (\ref{distortion-rate}) and (\ref{cap-region}), the problem can be formulated as
%%%%%%%%%%%%%%%%%%%%%%%%%%%%%%%%%%%%%%%%%%%%%%%%%%%
\begin{eqnarray} \label{mb}
\mbox{minimize} & & b \nonumber \\
\mbox{subject to} & & b \sum_{j=1}^l  R_j\geq \Rc(d_l), \;\;\; \forall \; l \nonumber \\
& & \sum_{l=1}^L \frac{R_l}{C_l} \leq 1, \;\;\; R_l \geq 0 , \;\; \forall \; l
\end{eqnarray}
%%%%%%%%%%%%%%%%%%%%%%%%%%%%%%%%%%%%%%%%%%%%%%%%%%%
Letting $\Rc(d_0) = 0$, we easily see that a necessary condition for optimality is that
the inequalities $b R_l \geq \Rc(d_l) - \Rc(d_{l-1})$ must hold with equality for all $l = 1,\ldots,L$.
Replacing these into the capacity region constraint, we obtain the solution
%%%%%%%%%%%%%%%%%%%%%%%%%%%%%%%%%%%%%%%%%%%
\begin{equation} \label{mb-sol}
b^\star = \sum_{l=1}^L \frac{\Rc(d_l) - \Rc(d_{l-1})}{C_l}.
\end{equation}
%%%%%%%%%%%%%%%%%%%%%%%%%%%%%%%%%%%%%%%%%%%%
{\bf Min Weighted Total Distortion} (MWTD): For given $b$ and non-negative weights $\{w_l\}$,
we wish to minimize $\sum_{l=1}^Lw_l D_l$.

In ~\cite{DCC-09}, the case $s=1$ is solved in closed form. The general case $s > 1$ treated here is
more difficult. However, we can cast MWTD as a convex optimization problem with respect to the variables
$\{R_l\}$ and $\{D_{i,l}\}$ that can be solved by standard tools
by incorporating the reverse waterfilling solution of the $\Rc(D)$ into the problem itself. We have
%%%%%%%%%%%%%%%%%%%%%%%%%%%%%%%%%%%%%%%%%%%%%%
\begin{eqnarray} \label{eq:MWD}
\mbox{minimize} & & \sum_{l=1}^L w_l \left(\frac{1}{s}\sum_{i=1}^s D_{i,l} \right)\nonumber\\
\mbox{subject to} & & \sum_{l=1}^L \frac{R_l}{C_l} \leq 1, \;\;\; R_l \geq 0 , \;\; \forall \; l \nonumber \\
& &  0 \leq D_{i,l}\leq \sigma_i^2, \;\;\; \forall \; i, l \nonumber\\
&& \frac{1}{s}\sum_{i=1}^s \frac{1}{2} \log_2\frac{\sigma_i^2}{D_{i,l}} \leq  b \sum_{j=1}^l R_j , \;\; \forall \; l
\end{eqnarray}
%%%%%%%%%%%%%%%%%%%%%%%%%%%%%%%%%%%%%%%%%%%%%%%%%
{\bf Min-Max Distortion Penalty} (MMDP): For given $b$, we wish to minimize $\max_{l\in{\cal L}} D_l/D_l^{\rm opt}$,
where $D_l^{\rm opt} = {\cal R}^{-1}(bC_l)$ is the individual R-D bound for user $l$.

Again, in ~\cite{DCC-09} we treated the case $s=1$ that can be cast as a simple linear program.
In the general case $s > 1$, we obtain the following convex problem with respect to the variables $\alpha$,
$\{R_l\}$ and $\{D_{i,l}\}$ :
\begin{eqnarray}
\label{eq:MMDP}
\mbox{minimize} & &  \alpha \nonumber \\
\mbox{subject to} & & \sum_{l=1}^L \frac{R_l}{C_l} \leq 1, \;\;\; R_l \geq 0 , \;\; \forall \; l \nonumber \\
& &  0 \leq D_{i,l}\leq \sigma_i^2, \;\;\; \forall \; i, l \nonumber\\
&& \frac{1}{s}\sum_{i=1}^s \frac{1}{2} \log_2\frac{\sigma_i^2}{D_{i,l}} \leq  b \sum_{j=1}^l R_j , \;\; \forall \; l  \nonumber \\
&& \sum_{i=1}^s{D_{i,l}} \leq \alpha D^{\rm opt}_l, \;\;\; \forall \;\; l
\end{eqnarray}
\begin{figure}[tpb]
\centerline{\includegraphics[width=9cm,height=5cm]{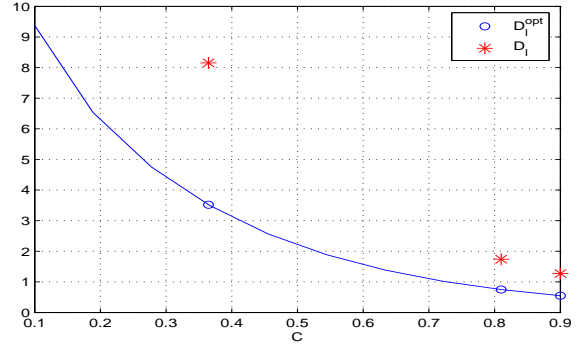}}
\caption{$D_l^{opt}$ for capacity values ranging from 0.1 to 0.9.  The $D_l$ point (stars) represent the solutions of the MMDP problem for the capacities  $\Cu=[0.3645,\;0.81,\;0.9]$ when 4 parallel sources with variances ${\underline {\sigma}}^2=[50,\;12,\; 8,\; 5]$ are considered. }
\label{fig:MDDP}
\vspace{-0.3 cm}
\end{figure}
In Fig. \ref{fig:MDDP}, we plot $D_l^{opt}$ for capacity values ranging from $0.1$ to
$0.9$. The points $D^l$ represent the solution of the MMDP
optimization problem for the capacities and the set of variances we use in Section \ref{sec:Results} for the simulations. For each considered capacity, the distortion penalty of each user can observed by comparing
the curve $D_l^{opt}$ with the distortion points $D_l$.

%%%%%%%%%%%%%%%%%%%%%%%%%%%%%%%%%%%%%%%%%%%%%
\section{Joint Source-Channel Coding for the BEBC} \label{sec:jscc}

In this section we present some code design examples that approach the theoretical limits obtained in
(\ref{mb-sol}), (\ref{eq:MWD}) and (\ref{eq:MMDP}).
Generalizing the scheme proposed in \cite{DCC-09}, the layered JSCC scheme for parallel Gaussian sources is
based on the concatenation of scalar embedded quantizers with linear encoding functions, implemented by
the same basic raptor encoding ``machine''.  We use an embedded scalar quantizer $\Qc : \RR \rightarrow \{0,1\}^N$, optimized
for a Gaussian source $\sim \Nc(0,1)$. In order to quantize the $i$-th source component,
the source symbols are scaled by $1/\sigma_i$ and the inverse scaling of the quantizer reconstruction points is applied at the decoder.

Let $\Um^{(i)} = \Qc(\Sm_{i,:}/\sigma_i)$ denote the sequence of quantization indices, formatted as an $N \times k$ binary array.
The $j$-th row of $\Um^{(i)}$, denoted by $\Um_{j,:}^{(i)}$, is referred to as the the $j^{\rm th}$ ``bit-plane''.
Without loss of generality, we let $\Um_{1,:}^{(i)}$ denote the sign bit-plane, and $\Um_{2,:}^{(i)}, \ldots, \Um_{N,:}^{(i)}$
denote the magnitude bit-planes with decreasing order of significance.
The quantizer output $\Um^{(i)}$ forms a discrete memoryless source, with entropy rate
$H^{(i)} = \frac{1}{k} H(\Um^{(i)})$ (in bits/source symbol). This can be decomposed as
$H^{(i)} = \sum_{j=1}^N H_j^{(i)}$, where the conditional entropy rates of the bit-planes are denoted by
$H_j^{(i)} = \frac{1}{k} H( \Um_{j,:}^{(i)} | \Um_{1,:}^{(i)},\ldots, \Um_{j-1,:}^{(i)})$, for $j = 1,\ldots,N$.
The R-D function achieved by embedded scalar quantization followed by entropy coding of the quantization indices
is given by the set of R-D points with coordinates $\left ( \sum_{j=1}^p H_j^{(i)} , D_{\Qc,i}(p) \right )$, for $p = 1,\ldots,N$,
where the quantizer distortion can be approximated by the following function
$D_{\Qc,i}(r) \approx \Gamma  \sigma_i^2\, 2^{-2r}$,
where $\Gamma$ is a multiplicative penalty factor \cite{Gish-Pierce, Ziv} and for the $i^{\text th}$ source $r=\sum_{j=1}^{p}H_j^{(i)}$ is the total rate of $p$ bit-planes.

\begin{figure}[tpb]
\centerline{\includegraphics[width=9 cm]{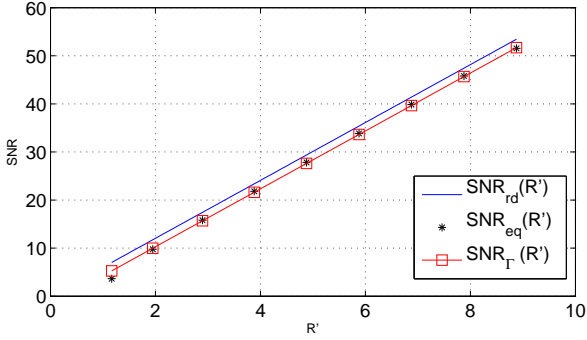}}
\caption{$\SNR$ vs. Rate, for a Gaussian source with $\sigma=1$, obtained by the R-D function ($\SNR_{rd}(R')$), by the set of R-D points of the embedded quantizer with coordinates $\left (\sum_{i=1}^p H_i , D_\Qc(p) \right )$ for $p = 1,\ldots,10$ ($\SNR_{eq}(R')$) and by $D_{\cal Q}(R')=\Gamma2^{-2R'}$ where $\Gamma=1.5$, ($\SNR_{\Gamma}(R')$).}
\label{fig:Gamma}
\vspace{-0.3 cm}
\end{figure}
We define reconstruction signal-to-noise-ratio SNR$\eqdef 10\log_{10}\frac{\sigma^2}{D}$.The SNR-R curve of the embedded scalar quantizer used in this work is shown in
Fig. \ref{fig:Gamma} for $\sigma=1$.
We notice that the familiar 6dB per quantization bit improvement in the reconstruction SNR is achieved
within a gap  $\Gamma \approx 1.5$.
%{\RED NOTE: I used $\Gamma$ instead of $\alpha$ since the latter was used before to denote the min-%max gap in the MMDP problem.}

The code design proceeds as follows: first, we approximate  the solution of the desired optimization problem (e.g., MB, MWTP, MMDP said above) by taking into account that the actual quantizer R-D function has gap $\Gamma$ and that
the source coding rates must be quantized into at most $N$ discrete values, reflecting the bit-planes.
The details of this approximation shall be given later on. For the time being,
it is sufficient to say that the result of this first design step is a bit-plane allocation,
defined by the integers $\{p_{i,l}\}$, such that $p_{i,l}$ is the number of bit-planes of the $i$-th source component
that must be used for source reconstruction  at the $l$-th decoder (we define $p_{i,0} = 0$ for notation convenience).

Then, for each $l$, we form the $l$-th layer channel codeword by collecting all the quantization
bit-planes with indices $p_{i,l-1} < j \leq p_{i,l}$ for all source components $i = 1,\ldots, s$,
and mapping them independently onto binary codeword $\xv_{i,j}$ via a linear raptor encoder.
Building on our results in~\cite{Ozgun-Maria-JSCC-08}, we use a {\em systematic binary raptor encoder} such that the source bit-planes form the systematic part,
and the produced codeword corresponds to the ``parity'' symbols (see \cite{shokr2006IT,IT_Fresia_Caire, Ozgun-Maria-JSCC-08, DCC-09} for details). We
use the following degree distribution \cite{shokr2006IT}:
\begin{eqnarray}\nonumber
\Omega(x) &\!\!\!\!\!\! =\!\!\!\!\!\! & 0.008x\! +\! 0.494x^2\! + \!0.166x^3\! +\! 0.073x^4\! +\! 0.083x^5 \\\nonumber
& &\!\!\!\!\! + \!0.056x^8\!+\!0.037x^9\!+\!0.056x^{19}\!+\! 0.025x^{65} \!+\! 0.003x^{66}\!.
\end{eqnarray}
Finally, the $\sum_{i=1}^s p_{i,L}$ codewords $\xv_{1,1},\ldots,\xv_{s,p_{s,L}}$
are transmitted in sequence. This corresponds to the time-sharing nature of the
BEBC capacity region. We note here that the very same scheme can be applied with minor modifications to any underlying
$L$-user broadcast channel, if one is willing to accept the performance penalty of using time-sharing instead
of an optimal broadcast coding strategy.

At each $l^{\rm{th}}$ decoder,  all codewords from $(i,1)$ to $(i,p_{i,l})$ for all source components $i = 1,\ldots, s$
are sequentially decoded using the successive multi-stage decoder described in \cite{Ozgun-Maria-JSCC-08, IT_Fresia_Caire}. Each $(i,j)$-th stage of the multistage decoder is an iterative Belief Propagation decoder that incorporates
the information corresponding to the a priori probability distribution of the bit-plane symbols
and the posterior log-likelihood ratios produced by the decoders at previous stages,
$(i,1) \ldots, (i,j-1)$.
We omit the detailed description of the decoder because of space limitation.
If the code is well designed, the distortion achieved at decoder $l$ is essentially equal to the quantization distortion,
given by $ D_l \approx \small{\frac{\Gamma}{s}} \sum_{i=1}^s  \sigma_i^2\, 2^{-2\sum_{j=1}^{p_{i,l}}H_j^{(i)}}$.

In the rest of this section we discuss the details of the bit-plane allocation to the layers and of the choice of the coding rates.
Let $\{D^\star_{i,l}\}$ denote the set of distortions resulting from the solution of one of the optimization problems of Section \ref{sec:optimization}. We choose $p_{i,l}$ such that $\Gamma\sigma_i^22^{-2\sum_{j=1}^{p_{i,l}}H_j^{(i)}} \approx {D^\star_{i,l}}$.

 Since $p_{i,l}$ must be an integer, this step requires some ad-hoc approximation.
Recalling that the bit-planes with indices $p_{i,l-1} < j \leq p_{i,l}$ must be decoded by user $l$,
the block length of codeword $\xv_{i,j}$ is given by
\begin{equation} \label{block-length}
n_{i,j} = \frac{k H_j^{(i)}}{C_l - \delta_l}
\end{equation}
where $\delta_l$ denotes a suitable ``gap-to-capacity'' (or ``overhead'') of the raptor code
(see \cite{shokr2006IT}). The overall bandwidth expansion factor of the scheme is given by
\begin{equation}\nonumber  \label{bw-effective}
\widehat{b}  = \frac{1}{sk} \sum_{l=1}^L \sum_{i=1}^s \sum_{j = p_{i,l-1}+1}^{p_{i,l}} \!\!\!n_{i,j} = \frac{1}{s}\sum_{l=1}^L \sum_{i=1}^s \sum_{j = p_{i,l-1}+1}^{p_{i,l}} \frac{H_j^{(i)}}{C_l - \delta_l}.
\end{equation}
%\begin{eqnarray} \label{bw-effective}
%\widehat{b} & =& \frac{1}{sk} \sum_{l=1}^L \sum_{i=1}^s \sum_{j = p_{i,l-1}+1}^{p_{i,l}} n_{i,j} \nonumber \\
%& = & \frac{1}{s}\sum_{l=1}^L \sum_{i=1}^s \sum_{j = p_{i,l-1}+1}^{p_{i,l}} \frac{H_j^{(i)}}{C_l - \delta_l}.
%\end{eqnarray}

%%%%%%%%%%%%%%%%%%%%%%%%%%%%%%%%%%%%%%%%%%%%%%%%%%%%
\vspace{-1mm}\section{Results} \label{sec:Results}

In this section, we report a few simulation results for source block length $k=10000$, with $L=3$ users
with capacities $\Cu=[0.3645,\;0.81,\;0.9]$ and $s = 4$ source components with variances
${\underline {\sigma}}^2=[50, \;12,\; 8,\; 5]$.

For the MMDP optimization scenario, we fix $b=2.5$.
The optimal values of $D_{i,l}^*$ obtained by solving (\ref{eq:MMDP}) are given in Tab.~\ref{table:MMDR} (top). %We target at possible $D_{i,l}$ values provided by scalar quantizer closest to optimal $D_{i,l}^*$.
%{\RED Question: are these values obtained by including the scalar quantizer penalty factor $\Gamma$
%in the formulation of the problem, or these values are the actual solution of the problem and then you have multiplied by $\Gamma$ in order to do the quantization? Have we addressed this point somewhere?}
The allocation of bit-planes to layers are done as described earlier. The
resulting $D_{i,l}$ values and the bit-plane allocation variables $p_{i,l}$'s are reported in Tab.~\ref{table:MMDR} (bottom).
As a sanity check, if we use the values $D_{i,l}$ of Tab.~\ref{table:MMDR} (bottom) to obtain target distortions $d_l$'s in
the MB optimization problem, we find that the required bandwidth expansion factor is  $b_{min}=2.58$.
Hence, the rounding of the optimal distortion values has only a marginal effect on the overall theoretical bandwidth efficiency.
The actual linear coding and soft-bit reconstruction JSCC scheme, after careful optimization of
the raptor code redundancy overhead, achieves the desired distortions at ${\hat b}_{JSCC} = 3.59$.
Similar results are reported for the MWTD optimization scenario, where we fix $b = 8.75$.
The solution of (\ref{eq:MWD}) is reported in Tab.~\ref{table:MWD} (top) and the actual
code performance, obtained by simulation, is given in Tab.~\ref{table:MWD} (bottom).
%
%The resulting optimal values of $D_{i,l}^*$ and $\SNR_{i,l}^*$ are provided in Tab.~\ref{table:MWD_opt}. The allocation of bitplanes to layers are done as described earlier. The resultant $\SNR_{i,l}$ values and the bitplane allocation variables $p_{i,l}$'s can be seen in Tab.~\ref{table:MWD_sim}. Again using $\SNR_{i,l}$ values of Tab.~\ref{table:MMDR_sim} in MB optimization problem we found minimum value to be $b_{min}=8.83$. According to our simulations, JSCC scheme required a ${\hat b}_{JSCC}=10.75$.
\small{

\begin{table}  \caption{MMDP, $b=2.5$}
  \centering
\begin{tabular}{|c|c|c|c|c|}
  \hline
  $b_{min}=2.58$ & $\Sm_{(1,:)}$ & $\Sm_{(2,:)}$ & $\Sm_{(3,:)}$ & $\Sm_{(4,:)}$ \\\hline
  $D^*_{:,1}$ & 9.8 & 9.8 & 8 & 5 \\
  $D^*_{:,2}$  & 1.74 & 1.74 & 1.74 & 1.74 \\
  $D^*_{:,3}$ & 1.27 & 1.27 & 1.27 & 1.27\\
  \hline
\end{tabular}

\vspace{.5 cm}
MMDP, Simulation Result
\vspace{0.2 cm}

\begin{tabular}{|c|c|c|c|c|}

  \hline
  ${\hat b}_{JSCC}= 3.59$ & $\Sm_{(1,:)}$ & $\Sm_{(2,:)}$ & $\Sm_{(3,:)}$ & $\Sm_{(4,:)}$ \\\hline
  $p_{:,1}$ & 3 & 0 & 0 & 0 \\
  $D_{:,1}$& 5.27 & 12 & 8 & 5 \\\hline
  $p_{:,2}$  & 4 & 3 & 2 & 2 \\
   $D_{:,2}$& 1.32 & 1.26 & 3.46 & 2.16 \\\hline
  $p_{:,3}$ & 4 & 3 & 3 & 2 \\
   $D_{:,3}$ & 1.32 & 1.26 &0.84 & 2.16 \\
  \hline
\end{tabular}
\label{table:MMDR}
\end{table}

\begin{table}  \caption{MWTD, $b=8.75$}
  \centering
\begin{tabular}{|c|c|c|c|c|}
  \hline
  $b_{min}=8.83$ & $\Sm_{(1,:)}$ & $\Sm_{(2,:)}$ & $\Sm_{(3,:)}$ & $\Sm_{(4,:)}$ \\\hline
  $D^*_{:,1}$ & 0.22 & 0.22 & 0.22 & 0.22 \\
  $D^*_{:,2}$  & 0.09 &0.09 & 0.09 & 0.09 \\
  $D^*_{:,3}$ & 0.09 &0.09 & 0.09 & 0.09 \\
  \hline
\end{tabular}

\vspace{.5 cm}
MWTD, Simulation Result
\vspace{0.2 cm}

\begin{tabular}{|c|c|c|c|c|}
  \hline
  ${\hat b}_{JSCC}=10.75$ & $\Sm_{(1,:)}$ & $\Sm_{(2,:)}$ & $\Sm_{(3,:)}$ & $\Sm_{(4,:)}$ \\\hline
  $p_{:,1}$ & 5 & 4 & 4 & 4 \\
  $D_{:,1}$& 0.33& 0.32 & 0.21 & 0.13 \\\hline
  $p_{:,2}$  & 6 & 5 & 5 & 4 \\
   $D_{:,2}$& 0.08 & 0.08 & 0.05 & 0.13 \\\hline
  $p_{:,3}$& 6 & 5 & 5 & 4 \\
   $D_{:,3}$ & 0.08 & 0.08 & 0.05 & 0.13 \\
  \hline
\end{tabular}
\label{table:MWD}
\end{table}

}
\bibliographystyle{IEEEtran}

\end{document}